\documentclass[prb,twocolumn,showpacs]{revtex4}
\usepackage[dvips]{graphicx}
\usepackage{epsfig}
\usepackage{psfrag}
\usepackage{amsmath}
\usepackage{amsfonts}
\usepackage{amssymb}
\usepackage{bm}
\begin{document}
\title{Bound whispering gallery modes in circular arrays of dielectric spherical particles}
\author{A. L. Burin$^{\dag\spadesuit}$}
\affiliation{$^{\dag}$Department of Chemistry, Tulane University, New Orleans, LA 70118}
\affiliation{$^{\spadesuit}$FTPI, School of Physics and Astronomy, University of Minnesota, 
Minneapolis, MN D-01187}
\date{\today}

\begin{abstract}
Low-dimensional ordered arrays of optical elements can possess bound modes having an extremely high quality factor. Typically, these arrays consist of metal elements which have significantly high light absorption thus restricting performance. In this paper we address the following question: can bound modes be formed in dielectric systems where the absorption of light is negligible? Our investigation of circular arrays of spherical particles shows that (1) high quality modes in an array of 10 or more particles can be attained at least for a refractive index $n_{r}>2$, so optical materials like TiO$_{2}$ or GaAs can be used; (2) the most bound modes have nearly transverse polarization perpendicular to the circular plane; (3) in a particularly  interesting case of TiO$_{2}$ particles (rutile phase, $n_{r}=2.7$), the quality factor of the most bound mode increases almost by an order of magnitude with the addition of $10$ extra particles, while for particles made of GaAs the quality factor increases by almost two orders of magnitude with the addition of ten extra particles. We hope that this preliminary study will stimulate experimental investigations of bound modes in  low-dimensional arrays of dielectric particles. 
\end{abstract}

\maketitle

{\bf 1.} Assemblies of microsize particles are useful in a variety of optical applications because of their resonant interaction with a visible light. Although most of the present research is focused on three dimensional ($3-d$) systems (e. g. photonic crystals), low-dimensional ($1-d$ and $2-d$) structures are attracting increasing attention because they can be more easily constructed and maintained than their $3-d$ counterparts. Despite its low dimension, $1-d$ structures have a remarkable array geometry effect on absorption, photoluminescence and Raman scattering of light.\cite{1-drev,George1,Terry1,VM} They can even serve as a $1-d$ nano-waveguide.\cite{MetalGuide} 

\bigskip

\begin{figure}[ptb]
\centering
\includegraphics[width=3in,clip]{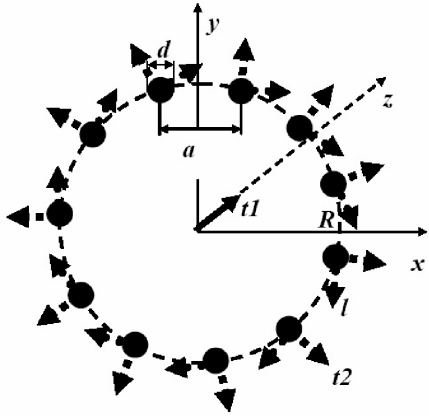}\caption{Circular array of particles. Arrows indicate polarizations of whispering gallery modes, $t1$ along the z-axis; $t2$ along the radii and $l$ tangential to the circle, both in the $x-y$ plane.}
\label{fig:pict}
\end{figure}

\bigskip

The strong dependence of light scattering and absorption on  $1-d$ array geometry suggests  that the quasi-bound optical modes exist in extended $1-d$ arrays. The resonant interaction of light with those modes is responsible for the optical  properties of arrays. Each mode can be characterized by the complex eigenfrequency $z=\omega - i\gamma/2$, with the real part $\omega$ representing the mode frequency and the imaginary part $\gamma$ representing the mode decay rate. \cite{B1} Modes of interest must have a large enough quality factor $Q = \omega/\gamma$ so that they are {\it bound} to the whole  1-d structure. 

Indeed, $1-d$ arrays of $N$ identical particles can possess bound modes with a quality factor approaching infinity for $N \rightarrow \infty$,\cite{book,AF,FW,B1,SY} so $\gamma(N=\infty)=0$. The case $Q \rightarrow \infty$ takes place in a linear chain of identical particles separated by a small distance $a$ that is less than half of the resonant wavelength $\lambda$ 
\begin{equation}
a<\lambda/2=c/(2\omega). 
\label{eq:light_c2}
\end{equation} 
If Eq. (\ref{eq:light_c2}) is satisfied, then there exists at least one quasi-bound mode having a  maximum wavevector $q_{max}=\pi/a$ exceeding the wavevector of the resonant photon $k=2\pi/\lambda$. When the chain is infinite the decay of such mode is forbidden by the momentum conservation law requiring $k\geq q$. In the finite system of $N$ particles, the decay rate is finite, but it tends to zero when $N \rightarrow \infty$ in accordance with the power law $\gamma(N) \propto N^{-3}$ (see Ref. \cite{B1}). 

Quasi-bound modes of the finite system  possess a  very narrow frequency resonance that can be used in a variety of photonics applications including waveguides,\cite{MetalGuide} antennas and detectors.\cite{book,AF,SY} These applications involve the emission or absorption of light in a very narrow frequency range, and the capability of functioning in  the single photon regime.\cite{singlephoton} This narrow resonance leads to a well resolved far field emission pattern that can be used as a guiding signal for aircraft.\cite{KOW}  In addition, systems possessing modes with a high quality factor can be used in lasers because these modes have a low pump threshold for lasing. \cite{Lawandy,CaoVardeny,Lagendijk,Genack}

{\bf 2.} One can expect that the highest quality factor can be attained for whispering gallery modes in an array of particles arranged in a circle and equidistant from each other (Fig. \ref{fig:pict}). A circle  has no sharp ends in contrast with a particle chain where the lifetime of a mode is limited to the travel time between the ends of the chain. \cite{B1} 
Indeed, it has been demonstrated that whispering gallery modes in a circular array of cylinder-shaped antennas perpendicular to the circle plane possess a quality factor that grows exponentially as the number of antennas increases. \cite{FW,book,R,2dcalc}) Similarly, as described in Ref. \cite{R}, an exceptionally high quality factor of circularly shaped structural fluctuations was suggested to account for lasing in random media. Note that these modes can also produce a large Raman surface enhancement.\cite{George2}

\begin{figure}[ptb]
\centering
\includegraphics[width=3in,clip]{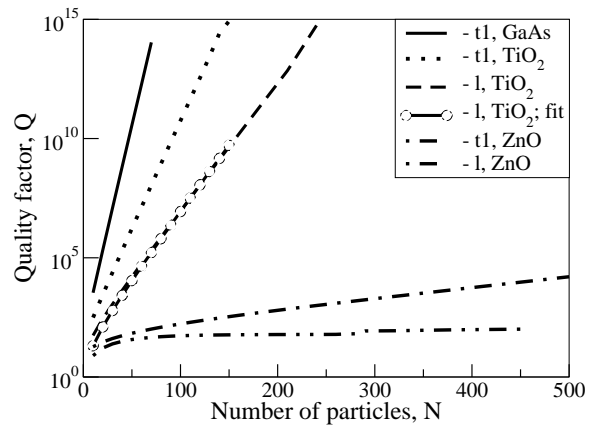}
\caption{Quality factor of $q=\pi/a$ modes versus the number of particles for $t1$, and $l$ modes in TiO$_{2}$
(rutile phase, $n_{r}=2.7$) and ZnO and $t1$ mode in GaAs ($n_{r}=3.5$). The fit of the quality factor by an inverse Bessel
function Eq. (\ref{eq:Bess_est}) $Q=0.2598/\mid J_{kaN/(2\pi)}\mid^{2}$ is shown for the longitudinal mode in TiO$_{2}$ (the product $qa$ has been defined by the solution of Eq. (\ref{eq:l}) at $N=110$. The mode $t2$ is not shown because it is hardly distinguishable with the mode $t1$ at $N \gg 1$.}
\label{fig:bessel}
\end{figure}

We believe that the exponentially small decay rate of whispering gallery modes is a common property of appropriately constructed circular arrays. This expectation can be justified as follows. Consider an array of $N$ identical particles forming a circle of radius $R$  and separated by distance $a$ (see Fig. \ref{fig:pict}). The symmetry of the problem suggests that rotation by $2\pi/N$ radians will change eigenmode amplitudes by a factor of $exp(iqa)$, where the wavevector $q=m/R$ ($m= 0, -1, 1, -2, 2, .. N/2$), enumerates $N$ eigenmodes possessing different quasi-angular momenta $m$. We will consider only even numbers of $N$ for the sake of simplicity. The mode can decay by the emission of a photon. The angular momentum of the emitted photon takes on the discrete set of values $m$, $m+N$, $m-N$, ... due to  rotational symmetry. We are primarily interested in the most efficient regime, when the angular momentum is equal to its minimum value $m$. The rate of emission of a photon possessing angular  momentum $m$ is defined by the squared overlap integral of the bound guiding mode and the photon wavefunction. This integral can be estimated using the wavefunction of a photon within the circular array (Fig. \ref{fig:pict}) positioned as characterized by the cylinder coordinate $\rho = R$. This wavefunction is given by the Bessel function $J_{m}(kR)$, where $k=\omega/c$ is the wavevector of an emitted photon, and $\omega$ is its frequency. Thus one can estimate the whispering gallery mode decay rate dependence on the number of particles $N$ in the circular array as 
\begin{equation}
\gamma(k) \propto |J_{qR}(kR)|^{2}, N=2\pi R/a. 
\label{eq:Bess_est}
\end{equation} 
This result reproduces the estimate of Ref. \cite{FW} for interacting antennas in the very similar case of a mode possessing a certain quasi-angular momentum. 
If the argument of the Bessel function is smaller than its index 
($k < q$ in accordance with Eq. (\ref{eq:light_c2})), then using the limit $N \rightarrow \infty$ one can approximate Eq. (\ref{eq:Bess_est}) by the exponential function 
\cite{KOW} (see Fig. \ref {fig:bessel}) 
\begin{eqnarray}
\gamma \propto exp(-\kappa aN/\pi), 
\label{eq:Bess_est_1}\\
\kappa=q (cosh^{-1}(1/x) - \sqrt{1-x^2}), ~ x=k/q.
\nonumber
\end{eqnarray}

\bigskip

\bigskip

Thus our estimate shows that the radiative decay rate of the whispering gallery mode can be made exponentially small. Exponentially narrow resonances (Eq. (\ref{eq:Bess_est_1})) can contribute to the optical properties of the system only when the width of the frequency resonance  (Eq. (\ref{eq:Bess_est})) exceeds the rate of light absorption by particles. Metal particles or antennas always have a significant absorption due to conducting electrons. This is not the case for dielectric particles. They  have negligible absorption when the photon energy is less than the width of the forbidden band. In addition, the optical (Mie) resonance frequency of a dielectric particle is inversely proportional to its size and can be easily changed to the desired value, while the resonance of metal particles is bound to their plasma frequency at least until the particle size is less than the wavelength. Therefore we find it important to test whether narrow resonances (Eq. (\ref{eq:Bess_est})) can be attained in the array of dielectric particles possessing a realistic refractive index $n_{r}$.   

{\bf 3.} Bound modes have to satisfy Eq. (\ref{eq:light_c2}). Formally, this goal can be attained by reducing the interparticle distance $a$. However, the interparticle distance cannot be made smaller than the spherical particle diameter $d$, so we cannot satisfy the condition $k<q$ at any refractive index. Since the resonant wavelength increases with increasing refractive index $n_{r}$ approximately as $\lambda \propto n_{r}$, bound modes will definitely exist at sufficiently large $n_{r}$.   Standard optical materials have refractive indices of order of $1$ (see Table \ref{tab:1}). It is not clear whether arrays of particles made of those materials can possess bound modes. Below, we employ the multiple sphere Mie scattering formalism \cite{classics} (see also \cite{Alexei1,An}) generalized  to the eigenmode problem, and show that  bound longitudinal modes exist at least for particles made of GaAs and TiO$_{2}$ and for any other material with the refractive index $n_{r}>1.9$. The bound transverse modes should exist at arbitrary refractive index at sufficiently large number of particles as one can expect in analogy with the weakly guiding optical fibers.\cite{books,comment_transv}

Eigenmodes of the array can be found by solving corresponding Maxwell equations without incident field. One can express electric and magnetic fields at a given frequency $z$ (wavevector $k=z/c$) by expanding them over spherical vector functions weighted with scattering amplitudes $a^{l}_{mn}$, $b^{l}_{mn}$ for each sphere $l$ ($l=1, 2, 3, ... N$),
where index $n= 1, ..$ stands for the photon angular momentum and index $m=-n, -n+1, .. n$ is the angular momentum projection onto the $z$-axis.\cite{classics} Eigenmode amplitudes $a$ and $b$ are bound by the set of   equations 
\begin{eqnarray}
0=\frac{a^{l}_{mn}}{\overline{a_{n}^{l}}}+(\widehat{A}a)^{l}_{mn}+(\widehat{B}b)^{l}_{mn},
\nonumber\\
0=\frac{b^{l}_{mn}}{\overline{b_{n}^{l}}}+(\widehat{B}a)^{l}_{mn}+(\widehat{A}b)^{l}_{mn}, 
\label{eq:hom1}
\end{eqnarray} 
where coefficients $\overline{a_{n}^{l}}$, $\overline{b_{n}^{l}}$ are the Mie scattering coefficients of the $l^{th}$ sphere.\cite{classics} In particular the dipolar scattering coefficients $\overline{a_{1}}$, $\overline{b_{1}}$ are defined as \cite{BookClassic}
\begin{eqnarray}
\overline{a_{1}}=\frac{\psi_{1}(kd/2)\psi_{1}'(n_{r}kd/2)-n_{r}\psi_{1}(n_{r}kd/2)\psi_{1}'(kd/2)}{\varsigma_{1}(kd/2)\psi_{1}'(n_{r}kd/2)-n_{r}\psi_{1}(n_{r}kd/2)\varsigma_{1}'(kd/2)};
\label{eq:Mie1}
\\
\overline{b_{1}}=\frac{n_{r}\psi_{1}(kd/2)\psi_{1}'(n_{r}kd/2)-\psi_{1}(n_{r}kd/2)\psi_{1}'(kd/2)}{n_{r}\varsigma_{1}(kd/2)\psi_{1}'(n_{r}kd/2)-\psi_{1}(n_{r}kd/2)\varsigma_{1}'(kd/2)};
\label{eq:Mie2}
\\
\psi_{1}(x)=sin(x)/x-cos(x), \varsigma_{1}(x)=e^{ix}(-1-i/x).
\nonumber
\end{eqnarray}

Matrices $\widehat{A}$ and $\widehat{B}$ define the interaction of multipole polarizations of different spheres. 
Their matrix elements have a general structure $M^{jl}_{mn\mu\nu}=exp(i(m-n)\phi_{lj})\sum_{p}D_{p}(\theta_{lj})exp(ikr_{lj})/r_{lj}^{p}$, where $r_{lj}$, $\phi_{lj}$, $\theta_{lj}$ are the spherical coordinates of the center of the sphere $j$ with respect to the center of the sphere $l$; $k=z/c$; and $D_{p}(\theta)$ are the real coefficients  as defined in Ref. \cite{classics}. The solution of Eq. (\ref{eq:hom1}) exists for the discrete set of frequencies $z_{a}=\omega_{a}+i\gamma_{a}$, $a=1, 2, 3 ..$, where these frequencies are defined by  eigenfrequencies of modes $\omega_{a}$ and their decay rates $\gamma_{a}$. 

The rotational symmetry of the problem (Fig. \ref{fig:pict}) permits us to seek solutions of Eq. (\ref{eq:hom1}) in the form 
\begin{eqnarray}
a^{l}_{mn}(q)= a_{mn}e^{2\pi i (m+qa)l}, b^{l}_{mn}=b_{mn} e^{2\pi i (m+qa)l},
\label{eq:sol_form}
\end{eqnarray}
with wavevector $q=2\pi p/(Na)$ ($p=-1, 1, -2, 2 ...N/2$). 
The dimension of Eq. (\ref{eq:hom1}) is reduced by a factor of $N$, because the intersphere interactions $\widehat{A}$ and $\widehat{B}$ get replaced with their Fourier transforms. At the next step one should restrict the maximum value of the angular momentum to $n<n_{max}$ and solve the equations requiring the weak sensitivity of the solution to the increase of $n_{max}$. In this paper, we will study the simplest case $n_{max} =1$ for coefficients $b_{mn}$, while we set all coefficients $a_{mn}$ equal to zero. This approach is similar to that of the coupled dipolar approach. It is well justified for a large refractive index $n_{r}\gg 1$ and for modes corresponding to dipolar Mie resonances  possessing the lowest frequency.\cite{SY} In fact, we are interested in these modes  because they are most easily bound (cf. Eq. (\ref{eq:light_c2})). 

Optical materials like TiO$_{2}$  or ZnO (see Table \ref{tab:1}) do not possess a high enough refractive index to use this dipolar approach. However, bound modes as found in the dipolar approach will remain bound as higher multipoles are taken into account. In fact, higher multipoles add additional modes as generated by corresponding Mie resonances, and all are located above the dipolar resonance. Those resonances can shift the energy of dipolar resonance downwards only because of energy level repulsion. Such behavior has been seen, for instance, in Ref. \cite{classics2}, where the rigorous  approach is compared with the discrete dipolar approach. The approximating method underestimates the size of the sphere at the lowest Mie resonance, which is equivalent to overestimating the resonant frequency. Therefore Eq. (\ref{eq:light_c2}) will be better satisfied in the more rigorous approach because of the increase of resonant wavelength due to the reduction of resonant frequency. Thus our method gives upper estimates for the decay rate of bound modes and for the minimum refractive index $n_{r}^{*}$ necessary to form strongly bound modes.

\bigskip

\begin{figure}[ptb]
\centering
\includegraphics[width=3in,clip]{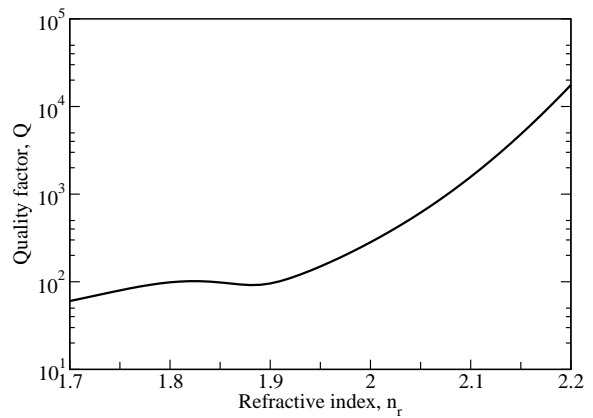}
\caption{Dependence of the quality factor of longitudinal mode on the refractive index near its threshold value $n_{r}^{*}\approx 1.9$ estimated for $N=100$ of dielectric spheres within the array}
\label{fig:threshold}
\end{figure}

\bigskip

In the dipolar approach, one can separate whispering gallery modes into three branches $t1$, $t2$ and $l$ depending on their polarization (Fig. \ref{fig:pict}). The particle polarization for the transverse mode $t1$ is perpendicular to the array on the $x - y$ plane and is described by the amplitude $a_{01}$ ($m=0$). The transverse mode $t2$ is characterized by particle polarizations close to the radial directions of the $x-y$ plane. It is defined by amplitudes 
$a_{\pm 11}$ ($a_{-11}\approx 2a_{11}$ \cite{comment1}). The longitudinal mode $l$ is characterized by the polarization that is approximately tangential to the circle and is defined by amplitudes 
$a^{l}_{\pm 11}$ ($a_{-11} \approx -2a_{11}$). In the case of most bound mode with the angular momentum $q=\pi/a$, the definitions of transverse and longitudinal modes are precise. 

\begin{table*}[t]
\begin{tabular}
[c]{|l|l|l|l|l|}\hline
Material & ZnO & TiO$_{2}$, anatase & TiO$_{2}$, rutile & GaAs \\\hline
$n_{r}$ & $1.7$ & $2.55$ & $2.7$ & $3.5$\\\hline
$\omega_{t}$ (s$^{-1}$) & $2.2\cdot 10^{15}$ & $1.8\cdot 10^{15}$ & $1.6\cdot 10^{15}$ & $1.2\cdot 10^{15}$ \\\hline
$\omega_{l}$ (s$^{-1}$) & $2.6\cdot 10^{15}$ & $1.9\cdot 10^{15}$ & $1.8\cdot 10^{15}$ & $1.4\cdot 10^{15}$\\\hline
Q, $N=10$, $t1$ & $12.7$ & $86.77$ & $124$ & $828$ \\\hline
Q, $N=10$, $l$ & $7.8$ & $40.3$ & $55$ & $303$ \\\hline
$dln(Q)/dN$, $t1$ & $0.0157$ & $0.175$ & $0.21$ & $0.4$\\\hline
$dln(Q)/dN$, $l$ & $0$ & $0.099$ & $0.131$ & $0.308$\\\hline
\end{tabular}
\caption{
Characterization of most bound modes in circular arrays of particles of $200$ nm diameter.  Mode parameters have been calculated for $N=10$ particles. 
\label{tab:1}}
\end{table*} 

The Fourier transform of interactions in the simplified Eq. (\ref{eq:hom1}) using the parameter definitions of Ref. \cite{classics} yields the dispersion equations for all three modes $\mu=t1$, $t2$ and $l$, given by 
\begin{equation}
\frac{i}{\overline{a}_{1}(z)}+\epsilon_{\mu}(k, q)=0, k=z/c,
\label{eq:gen_disp}
\end{equation}
characterized by the dispersion laws 
\begin{equation}
\epsilon_{t1}(k, q)=\frac{3}{2}\left(\Sigma_{1}(k, q)+i\Sigma_{2}(k, q)-\Sigma_{3}(k, q)\right), 
\label{eq:t1}
\end{equation}
\begin{eqnarray}
\epsilon_{t2}(k, q)=\frac{1}{4}(\epsilon_{+}(k, q) + \sqrt{\epsilon_{-}(k, q)^{2}+V(k, q)^2} 
\label{eq:t2}
\\
\epsilon_{l}(k, q)=\frac{1}{4}(\epsilon_{+}(k, q) - \sqrt{\epsilon_{-}(k, q)^{2}+V(k, q)^2} 
\label{eq:l}
\nonumber\\ 
\epsilon_{\pm}(k, q)=\frac{\sigma(k, q+2\pi/a)\pm\sigma(k, q-2\pi/a)}{2};
\nonumber\\
\sigma(k, q) = 0.75(\Sigma_{1}(k, q)-i\Sigma_{2}(k, q)+\Sigma_{3}(k, q)); 
\nonumber\\ 
V(k, q) = 0.75(\Sigma_{1}(k, q)+3i\Sigma_{2}(k, q)-3\Sigma_{3}(k, q)). 
\end{eqnarray}
where $\Sigma_{p}(q)=\sum_{l=1}^{N-1}exp(ikr_{0l}-iqal)/r_{0l}^{p}$, $r_{0l}=2Rsin(\pi l/N)$ is the distance between the centers of spheres separated by $l-1$ spheres (see Fig. \ref{fig:pict}) and $q_{\pm}=q\pm 2\pi/N$. For $N \rightarrow \infty$, the dispersions of modes $t1$ and $t2$ become identical because the circular array of large radius locally approaches a linear chain geometry where two transverse polarizations are equivalent. The ``photon'' field term $\Sigma_{1}(k, q)$ disappears in the $l-$mode spectrum Eq. (\ref{eq:l}) for $N\rightarrow \infty$ because photons  have a transverse polarization.  \cite{comment_transv}

{\bf 4.} One can resolve Eqs. (\ref{eq:t1}), (\ref{eq:t2}), (\ref{eq:l}) using the Newton-Raphson iteration algorithm for the equation 
\begin{equation}
f(k)=\frac{i}{\overline{a}_{1}(ck)}+\epsilon_{\mu}(k, q)=0  
\label{eq:gen_form}
\end{equation}
with the iteration procedure defined as
\begin{equation}
k_{n+1}=k_{n}-f(k_{n})/f'(k_{n}). 
\label{eq:NewtonRaphson}
\end{equation}
This procedure starts with some initial value $k=k_{0}$. At almost any initial value $k_{0}$, the iteration series rapidly converges to some solution of Eq. (\ref{eq:gen_form}). However, Eq. (\ref{eq:gen_form}) has generally infinitely many solutions and we are only interested in the solution possessing the largest quality factor. In particular, if there exists a quasi-bound mode possessing a quality factor much greater than unity, then the procedure should be organized in such a manner that it converges to this mode. In the regime of interest of Eq. (\ref{eq:light_c2}), where the bound mode exists, we have organized the iteration procedure choosing  the starting point $k_{0}=2/d$.  Using this choice, the algorithm converges to a frequency of the bound mode with the smallest decay rate in almost $100\%$ of our studies for $t1$ modes. The situation is more difficult for $t2$ and $l$ modes, where, at certain number of particles the algorithm Eq. (\ref{eq:NewtonRaphson}) does not converge being trapped by the attractor formed by the selfrepeating sequence of two or three different values of $q$ appearing periodically in Eq. (\ref{eq:NewtonRaphson}). In these cases the convergence has been reached by changing initial value of $q$ to $0.2/d$ or using the modified Newton Raphson algorithm $k_{n+1}=k_{n}-0.1f(k_{n})/f'(k_{n})$. When the quality factor is high ($Q \geq 20$) this is sufficient to find the right solution.    

In our calculations, we have used the particle diameter $d=2$ and took the minimum possible interparticle distance $a=d$, where bound modes can be formed most easily (Eq. (\ref{eq:light_c2})). Although the accurate analysis of electric and magnetic fields will be difficult in this case due to the singularity at the point where particles touch each other,\cite{Bruno} the dipolar approach should still serve as a good upper estimate for mode frequencies and decay rates as discussed above. The generalization to different particle sizes is straghtforward because characteristic frequences and wavectors behave as $q, k, z \propto 1/d$, while the quality factor is scale-invariant. The resonant frequencies for the particle size $d=200 nm$ for a large number of particles $N=10$ having different polarizations and refractive indices are given in Table \ref{tab:1} for the case of the most strong bound modes with $q=\pi/a$. We have also included mode quality factors for $N=10$ and their rates of exponential increase with the number of particles $N$. Fig. \ref{fig:bessel} illustrates the exponential dependence of the quality factor on the number of particles for different modes and materials.

Indeed, for large number of particles ($N>20$), the smallest decay rate has always been found at the maximum wavevector $q=\pi/a$ corresponding to the quasi-angular momentum $m=N/2$ (cf. \cite{AF,B1,SY} and we further discuss only this case). 
As transverse modes have the lowest frequencies $\omega_{l}(n_{r}, N)$, these modes are the most easily bound.  Bound  longitudinal modes are formed when  $\omega_{l}(n_{r}, q=\pi/a)<\omega_{thr}=c\pi/a$. According to our calculations, this happens at $n_{r}>n_{r}^{*}\approx 1.9$. The behavior of the quality factor of longitudinal mode  near the threshold is shown in Fig. \ref{fig:threshold}. For $n_{r}<1.9$ the quality factor for a circular array of $N=100$ dielectric particles depends very weakly on the refractive index, while the sharp, stepwise increase is clearly seen when the threshold is passed. Qualitatively similar behavior takes place at larger $N$. Also for $n_{r}<1.9$, the quality factor is almost independent of the number of particles (cf. Fig. \ref{fig:bessel}) while above this point the dependence is quite strong. As for transverse modes they can be made bound at arbitrary refractive index.\cite{comment_transv} However for $n_{r} < n_{*}$, a very large number of particles is required to attain a high quality factor (see Fig. \ref{fig:bessel} for ZnO).    

\bigskip

\begin{figure}[ptb]
\centering
\includegraphics[width=4in,clip]{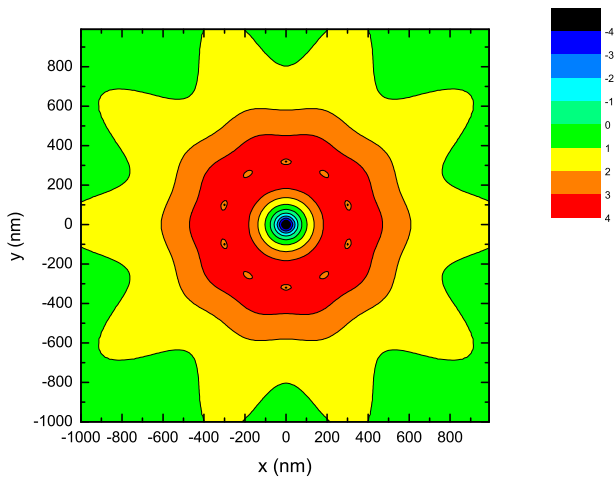}\caption{Contour plot describing the longitudinal whispering gallery mode intensity for $N=10$ GaAs particles ($n_{r}=3.5$) circular array. The intensity is expressed in arbitrarily units and contour lines separate domains for the intensity change by factor of ten (increase or decrease of the decimal logarithm of intensity by one. Particle sizes here and in Fig. \ref{fig:2C} are chosen $d=200nm$. }
\label{fig:2A}
\end{figure}

\begin{figure}[ptb]
\centering
\includegraphics[width=4in,clip]{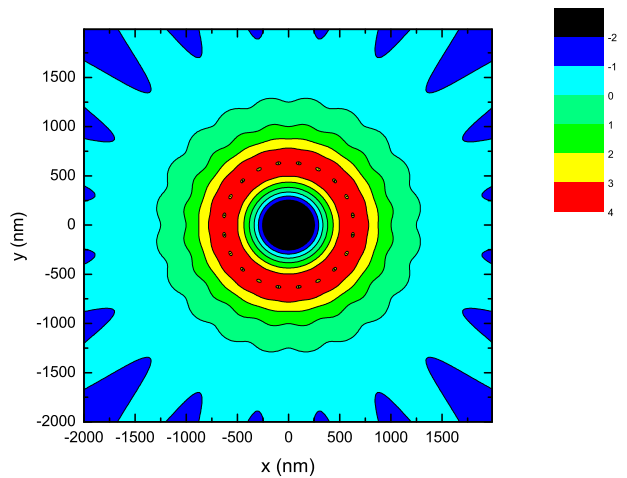}\caption{Contour plot describing the longitudinal whispering gallery mode intensity for $N=20$ GaAs particles ($n_{r}=3.5$).}
\label{fig:2C}
\end{figure}

\bigskip

{\bf 5.} Thus we can conclude that strongly bound longitudinal and transverse whispering gallery modes exist in circular arrays of TiO$_{2}$ and GaAs particles because their refractive indices exceed the upper estimate of the threshold $n_{r}^{*}=1.9$ (see Fig. \ref{fig:threshold}). 
Quality factors of most bound modes at the maximum wavevector $q=\pi/a$ for ZnO, TiO$_{2}$ and GaAs are shown in Fig. \ref{fig:bessel}. They all are exponentially sensitive to the number $N$ of particles and one can relate this dependence to that of the inverse squared Bessel function Eq. (\ref{eq:Bess_est_1}) as illustrated in Fig. \ref{fig:bessel}  for TiO$_{2}$. 

To characterize the spatial structure of whispering gallery modes, we have created contour plots for the intensity distribution of bound longitudinal modes within the array $x-y$ plane for GaAs particles.  The most bound modes ($q=\pi/a$) are shown in Figs. \ref{fig:2A}, \ref{fig:2C} for $N=10$ and $N=20$ particles respectively. In both cases the mode energy is highly concentrated within the array and the  shift from the array domain by the distance comparable or less to the particle size leads to the reduction of the mode intensity by orders of magnitude. The localization of optical energy is clearly  stronger in the case $N=20$. Indeed, relatively small displacement by $300$nm from the domain of array in any direction leads to a reduction of intensity by the factor of $10^{4}$ (only factor of $10$ in the case of $N=10$). This behavior agrees with the sharp increase of the quality factor with the number of particles for GaAs (see Fig. \ref{fig:bessel}). The strong reduction of intensity near the center of the circle is the consequence of the large angular momentum of the mode under consideration  (cf. Eq. (\ref{eq:Bess_est}) in the limit $r\rightarrow 0$).

As was discussed, the actual frequencies of low energy modes are smaller than our estimate, so the decay rates should decrease faster with $N$ than predicted (see Eq. (\ref{eq:Bess_est_1})). Even within our approach, the decay rate of the longitudinal mode for $n_{r}=3.5$ (GaAs) decreases by almost two orders of magnitude with the addition of ten particles to the array. To attain a similar effect in a $3-d$ photonic crystal, it can require the addition of $10^{3}\sim 1,000$ particles, which is much more difficult. Therefore, we believe that remarkable progress in photonics technology can be made by using $1-d$ circular arrays of dielectric particles.

This work is supported by the Air Force Office of Scientific Research (Award no. FA 9550-06-1-0110),  and by the TAMS GL fund (account 211043) through Tulane University. 
I greatly appreciate the hospitality of Boris Shklovskii at the University of Minnesota, where I relocated due to the hurricane disaster in New Orleans and where the essential part of this work has been completed. I wish to acknowledge Drs. R. Shore,  A. Yaghjian and A. Nachman for help in  constructing the problem and Gail Blaustein for her great help with manuscript preparation.



\begin{thebibliography}{99}
\bibitem{comment_transv} 
The analysis of the dipolar approach for transverse modes (see Eq. (\ref{eq:t1})) shows that  in the case $n_{r}>1$ there always exists the solution with the vanishing decay rate in the limit of $N \rightarrow \infty$. This solution exists due to the logarithmic divergence of the Fourier transform of the photon field interaction between spheres $\Sigma_{1}(k, q) \approx \ln\left(\frac{1}{a(q-k)}\right)$ in the limit of $k\rightarrow q$ (see \cite{VM}). The straightforward analysis of the limiting case $n_{r} \rightarrow 1^{+}$ shows that the most bound solution both for $t1$ and $t2$ modes is formed at $q=\frac{\pi}{a}$ and $k \approx q-\frac{1}{a}\exp\left(-\frac{2\pi}{(n_{r}-1)(\pi^{2}-8)}\right)$. This mode has a high quality factor at a very large number of particles $N>N_{*}=\exp\left(\frac{2\pi}{(n_{r}-1)(\pi^{2}-8)}\right)$ and the rate of increase of the quality factor with the number of particles scales as $1/N_{*}$. Our solution is similar to the specific class of solutions for $1-d$ waveguides obtained in Ref. \cite{books}. Since the dipolar approach fails in the limit of $n_{r} \rightarrow 1$ the quantitative and even qualitative applicability of the above estimate is under the question. The longitudinal mode spectrum Eq. (\ref{eq:l}) does not contain a logarithmically diverging term  $\Sigma_{1}(k, q)$ so we do expect that the bound modes vanishes at finite $n_{r} < n_{r*}$ with the threshold index $n_{r*}>1$ similarly to the consideration of this work. Preliminary estimates shows that this conclusion about $l-$modes remains valid in the approach taking into account the multipole interactions.
\bibitem{comment1}
The asymmetry between projections $m=\pm 1$ is caused by the definition of the Legendre functions $P_{1}^{\pm 1}$. \cite{classics}
\end{thebibliography}
\end{document}